\documentclass[aps,prl, showkeys]{revtex4}

\usepackage{amssymb}
\usepackage{amsmath}
\usepackage{graphicx}
\usepackage{float}
\usepackage{multirow}

\begin{document}

\title{Trajectory analysis through entropy characterization over coded representation }

% Use letters for affiliations, numbers to show equal authorship (if applicable) and to indicate the corresponding author
\author{Roxana \surname{Pe\~na-Mendieta}}
\affiliation{Facultad de Matem\'atica, Universidad de la Habana, San L\'azaro y L, CP 10400, Cuba}

\author{Ernesto \surname{Estevez-Rams}}
\email{estevez@fisica.uh.cu}
\affiliation{Instituto de Ciencias y Tecnolog\'ias de Materiales, Universidad de la Habana, San L\'azaro y L, CP 10400 Habana, Cuba}

\author{Ania \surname{Mesa-Rodr\'iguez}}
\affiliation{Facultad de Matem\'atica, Universidad de la Habana, San L\'azaro y L, CP 10400, Cuba}

\author{Daniel \surname{Estevez-Moya}}
\affiliation{Facultad de F\'isica, Universidad de La Habana, San Lazaro y L. CP 10400. La Habana. Cuba.}

\author{Danays \surname{Kunka}}
\affiliation{Institute of Microstructure Technology (IMT), Karlsruhe Institute of Technology (KIT), Hermann-von-Helmholtz-Platz 1, 76344 Eggenstein-Leopoldshafen, Germany}

%
% Keywords are not mandatory, but authors are strongly encouraged to provide them. If provided, please include two to five keywords, separated by the pipe symbol, e.g.:
\keywords{trajectories $|$ entropy $|$ complexity $|$} 

\begin{abstract}
Any continuous curve in a higher dimensional space can be considered a trajectory that can be parameterized by a single variable, usually taken as time. It is well known that a continuous curve can have a fractional dimensionality, which can be estimated using already standard algorithms. However, characterizing a trajectory from an entropic perspective is far less developed. The search for such characterization leads us to use chain coding to discretize the description of a curve. Calculating the entropy density and entropy-related magnitudes from the resulting finite alphabet code becomes straightforward. In such a way, the entropy of a trajectory can be defined and used as an effective tool to assert creativity and pattern formation from a Shannon perspective. Applying the procedure to actual experimental physiological data and modelled trajectories of astronomical dynamics proved the robustness of the entropic characterization in a wealth of trajectories of different origins and the insight that can be gained from its use. 
\end{abstract}

%\dates{This manuscript was compiled on \today}
%\doi{\url{www.pnas.org/cgi/doi/10.1073/pnas.XXXXXXXXXX}}

\maketitle

\section{Introduction}

Any continuous curve embedded in an n-dimensional space, from a mathematical point of view, can be considered a trajectory if every coordinate depends, at least, on an additional variable usually taken as time. Trajectories are ubiquitous in any field of natural science and lie at the core of physics. Generically speaking, a dynamical system is the study of trajectories, taken to be the solution of some set of differential equations. In many cases, the underlying laws that govern the trajectory are unknown, and the only data available are the trajectories themselves under various control parameters or experimental conditions. In other cases, although the trajectory equations may be known, their solution is only numerical. We are faced again with numerical data which has to be reduced and analyzed to extract regularities and patterns, in order to understand the system behaviour or deduce meaningfully classification schemes, or predict the trajectory beyond the available data.

In biological systems, the mobility of bacteria's has been, for example, subject to much studies, as well as other microscopic entities such as espermatozoids under different environmental conditions and mechanical constrains \cite{Bressloff13,hofling13,shaebani14,morales20,morales22}; the wandering of animals and their trajectories have implications in ecology studies \cite{benhamou06,codling08,fagan13}; the prediction of trajectory of extreme climate events such as hurricanes  has a long history of research and is still a subject of investigation \cite{witt23}; the trajectories and stability of motions in objects subject to pair interaction potentials, such as cosmic objects lies at the historical foundation of Physics and still draws attention \cite{reiff22}; health conditions has been associated to different behaviors in human gait, and balance which can be reduced to a study of trajectories under experimental settings \cite{yogev05,franchignoni10,amboni13,santos16,montesino18,din19}. In any case, tracking and characterizing trajectories can be a daunting task and is usually essential to understand a wide range of phenomena. In no few cases, understanding is precisely about capturing the features of the trajectories and their implications \cite{codling08,reyes23}. 

In two dimensions, a shape curve can be considered as a particular case of a trajectory and has been described by several methods, of which the use of Fourier descriptors is one of the most powerful techniques \cite{persoon76,kuhl82}. The extension to curves embedded in higher dimensional space is straightforward. The Fourier decomposition of a curve results in a number of discrete coefficients from which the original trajectory, in principle, can be recovered \cite{zahn72}. The Fourier coefficients, taken as weights over the different frequencies,  can be used to characterize the complexity of trajectories, including their regular or chaotic nature \cite{powell79}.

The complexity of a trajectory could be understood in terms of entropy if a suitable definition that allows it to be effectively calculated for a given data could be used. Kolmogorov-Sinai entropy could be a suitable choice, and other entropy-related magnitudes could also be used for such characterization, such as effective measure complexity \cite{grassberger86,crutchfield03,crutchfield12}. This could allow us to build complexity maps and follow the evolution of trajectories statistically. There is a close relation between these entropic magnitudes and algorithmic complexity, as defined by Kolmogorov \cite{kolmogorov65,vitanyi93}. For the purpose at hand, the algorithmic complexity of a trajectory would mean the length of the shortest algorithm from which the trajectory could be reconstructed. The availability of Universal Turing Machines makes such measures meaningful regardless of the formal particularities of the algorithmic description. Algorithmic complexity also allows us to define a distance between trajectories in an informational sense, which is a robust measure in a number of applications \cite{li04}.

We address here how such an entropic description of a trajectory can be done in practical terms, building from a time series and discrete coding of a trajectory. Our key results are as follows: (i) We describe a procedure to code arbitrary trajectories as a time series by discretization and show how usual characteristics, such as fractal dimension, can be computed from the resulting code. (ii) We define trajectory entropy and related quantities over the coding and discuss its interpretation in terms of information generation and time correlation at all scales. (iii) We show how the defined magnitudes can be used for classification purposes in three examples: in the known H\'enon-Heiles model of motion to distinguish between regular and irregular trajectories, in the detection of Parkinson's disease in subjects from their gait, in the analysis of falling danger for subjects from their posture data.

Although it will be apparent that the dimensionality curse haunts the approach followed, as the alphabet size for the trajectory coding will scale as a power of the space dimension, it is still usable for two- and three-dimensions, with a $8$ and $26$ symbols alphabet, respectively, making it suitable for a broad number of applications in a wide range of fields.

\section{Trajectory analysis}

\subsection{Freeman coding}

A procedure for encoding arbitrary geometric configurations using a chain code was given first by Freeman for two-dimensional curves \cite{freeman61,freeman61a}. The coding needs to strike a balance in the chain code that faithfully preserves the trajectory information describing the shape and allows compact storage while facilitating any further processing \cite{freeman74}.

\begin{figure*}[!ht]
\centering
\includegraphics*[scale=0.6]{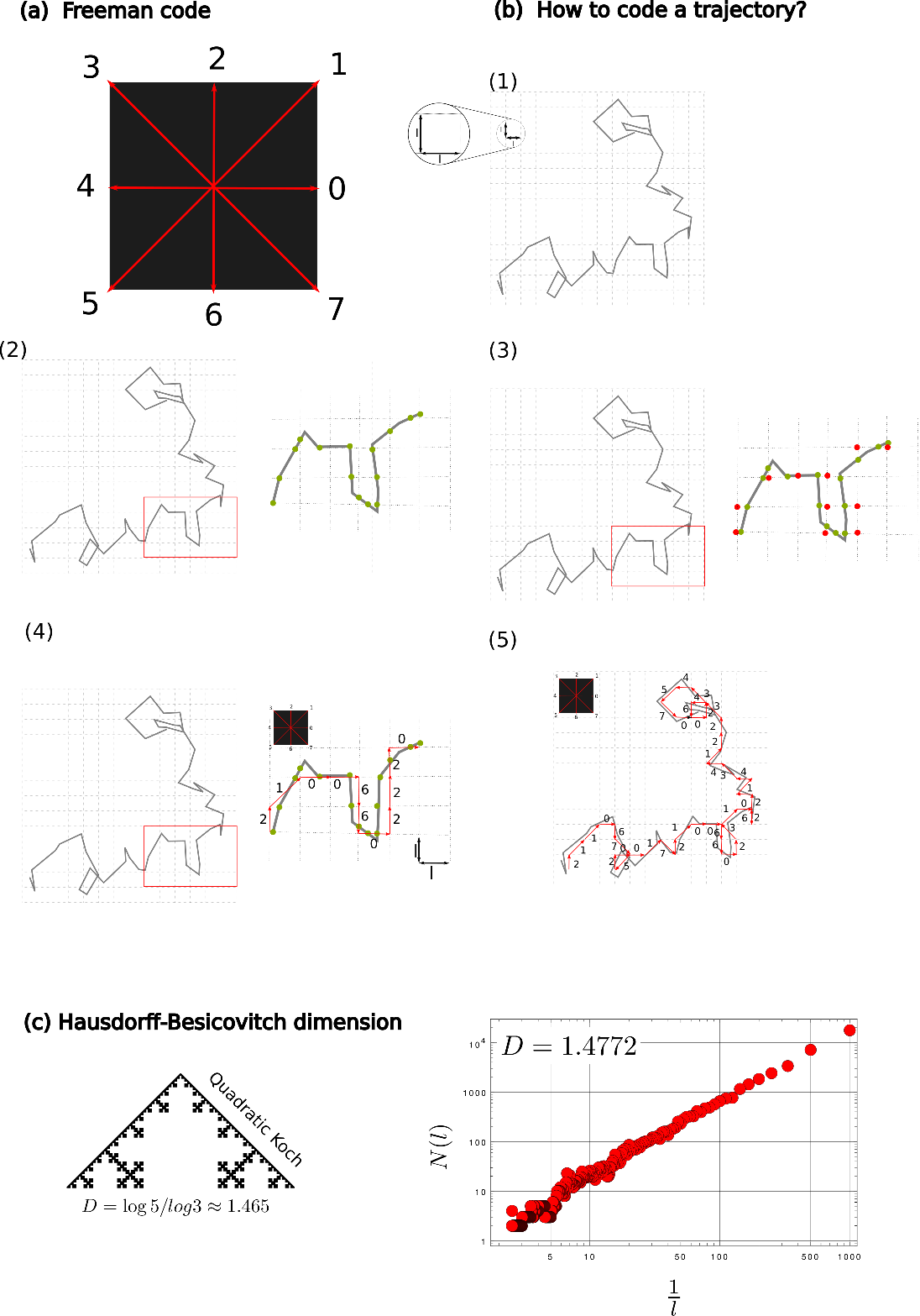}
\caption{\textbf{Trajectory encoding.} A trajectory can be encoded as a string using a finite alphabet $\chi$ such as the (a) Freeman alphabet in two dimensions. Freeman encoding (b) starts with imposing a square grid of length $l$ (1) over the trajectory. The intercept of the trajectory with the grid (2) determines the closest corner of the grid (3) to be taken as reference points; (4) from the reference points, the segments are determined, and characters from the alphabet area are assigned (See Supporting Information for code). The character string (5) follows. The length of the coded string for different grid sizes can be used to determine the Hasudoff-Bersicovitch (c) fractal dimension $D$. Following equation (\ref{eq:fractal}), the fractal dimension can be estimated from the slope of the log-log plot of the code length $N(l)$ vs $1/l$. The procedure was applied to a quadratic Koch curve where the fractal dimension is $\log 5/\log 3$. The estimated fractal dimension has an error of $0.8\%$ with respect to the true value. The fractal dimension of other fractal curves was also estimated with similar results.
}\label{fig:freeman}
\end{figure*}

The idea behind the Freeman encoding is to approximate the given trajectory by straight segments in any of the $8$ directions depicted in Figure \ref{fig:freeman}a. To build the segmented approximation, a square grid with edge length $l$ is superposed over the trajectory (Figure \ref{fig:freeman}b-1), and the intercept of the trajectory with the grid is determined (Figure \ref{fig:freeman}b-2). To each intercept, the closest vertex or node in the square grid is assigned (Figure \ref{fig:freeman}b-3), the straight segments are drawn between the chosen vertex, and an arrow points in the sense the trajectory is drawn (Figure \ref{fig:freeman}b-4). Finally, from the type of segment, in the sense of the trajectory, a chain code is built referred to the Freeman alphabet (Figure \ref{fig:freeman}b-5). 

Other encoding schemes have been discussed, with different grid shapes or how nodes are selected \cite{blumenkrans91,bribiesca99,bribiesca19}, but the one described will be sufficient for our analysis.

One first consideration is that the encoding allows the direct calculation of the fractal dimension of the trajectory \cite{mandelbrot67}. Richardson dimension estimation can be written as \cite{allen95}
\begin{equation}
 L(l)=Ml^{(1-D)},\label{eq:fractal}
\end{equation}
$L(l)$ is the length of a given curve measured using a yardstick of length $l$, and $D$ is the fractal dimension. 

In order to use the Freeman code for determining $L(l)$, one has to take into account that there are two lengths associated with the codes, $l$ for even codes $0,2,4,6$, and $\sqrt{2}l$ for odd codes $1,3,5,7$:
\begin{equation}
L(l)=\sum P[f(i)+1]l+P[f(i)]\sqrt{2}l,\label{eq:df}
\end{equation}
where $P[x]$ gives $0$ if $x$ is even, and $1$ otherwise. The sum is over the chain code $f_1f_2\ldots f_N$ of length $N$.

From equation [\ref{eq:fractal}], follows
\begin{equation}
 \log{\left [L(l) \right ]}=\log{M}+(1-D)\log{l}.\label{eq:fractal2}
\end{equation}
Taking $L(l)=l\times N(l)$,
 \begin{equation}
 \log{\left [N(l) \right ]}=\log{M}+D\log{\frac{1}{l}}.\label{eq:fractal1}
\end{equation}
A Richardson plot, $\log [N(l)] $ vs $\log[1/l]$ will then give the fractal dimension as the slope of the straight line as shown in Figure \ref{fig:freeman}c.

\subsection{Trajectory entropy and related magnitudes}

Kolmogorov-Sinai entropy rate, or entropy rate, $h$ will be used to measure information production \cite{cover06}. Given a bi-infinite sequence $F^{(t)}=\ldots f^{(t)}_{-2}f^{(t)}_{-1}f^{(t)}_{0}f^{(t)}_{1}f^{(t)}_{2}\ldots$, where $f^{(t)}_i$ is the symbol observed in the cell $i$ at time step $t$, KS-entropy density can be defined as \cite{crutchfield03}: 
\begin{equation}
 h=H(f_i|\ldots, f_{i-1}),\label{eq:ksent}
\end{equation}
where $H(X|Y)$ denotes the Shannon conditional entropy of random variable $X$ given variable $Y$ \cite{cover06}. Expression [\ref{eq:ksent}] defines the KS-entropy as the amount of new information in the observation of cell $f_i$, conditional on the state of all previous cells $f_j$, $j<i$. In this sense, the larger the value of $h$, the larger the randomness in the Freeman chain code.

For, necessarily, finite data, the KS-entropy has to be estimated \cite{schurmann99,lesne09}. In the methods section, we describe the Lempel-Ziv factorization \cite{lz76} procedure for such estimation.

A related magnitude, measuring the correlation at different scales, is the effective measure complexity $E$ \cite{grassberger86}, also known as excess entropy \cite{crutchfield03}.
\begin{equation}
\begin{array}{ll}
 E(S) &=I[\ldots, s_{-1}: s_0, s_1, \ldots],\label{eq:excess}
 \end{array}
\end{equation}
 $I[X:Y]=H[X]+H[Y]-H[X,Y]$ is the mutual information between $X$ and $Y$ \cite{cover06} and is a measure of the amount of information one variable carries regarding the other. $E$ is related to pattern production. For a periodic sequence of period $p$, $E=\log P$. The larger the amount of pattern in a sequence, the larger the value of $E$ and therefore measures correlations at all scales \cite{crutchfield03}. Effective measure complexity is estimated through Lempel-Ziv factorization using a random shuffle procedure \cite{melchert15}.

\subsection{Classification procedure}

\begin{figure}[!ht]
\centering
\includegraphics*[scale=0.8]{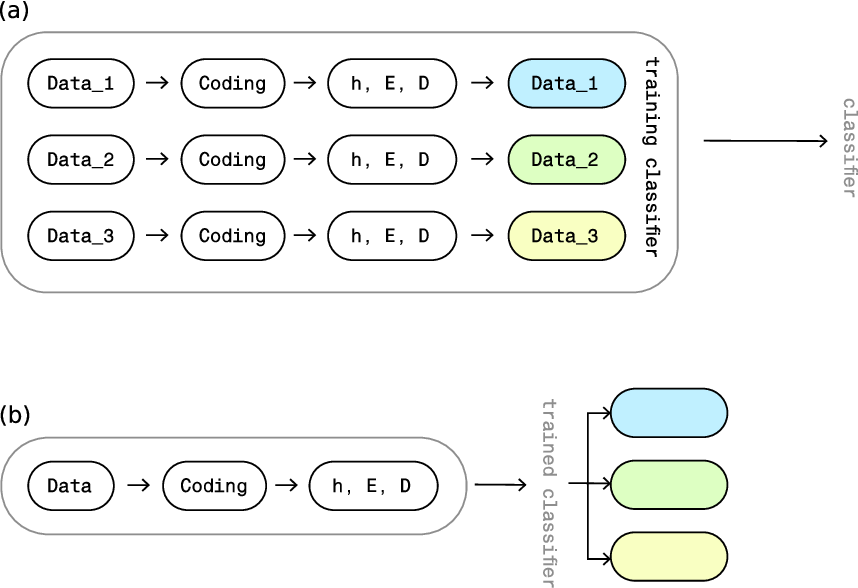}
\caption{\textbf{Classification scheme.} The trajectory data is discretized using Freeman coding. From the character stream, the entropy density ($h$), the effective measure complexity ($E$), and the fractal dimension ($D$) are estimated. (a) $(h, E, D)$ for each trajectory is fed into a machine learning (ML) procedure and the classification assignment in the training stage. (b) The trained ML is used for classifying other trajectories. Again, the trajectory is coded, and from the character stream, $(h,E,D)$ is estimated and fed into the classifier, which assigns a label to the data.
}\label{fig:classification}
\end{figure}

The details of the classification procedure are explained in the methods section. In brief, the classification starts with the Freeman encoding of the trajectories, from which the entropy density $h$, effective measure complexity $E$, and the fractal dimension are calculated. A set of trajectories is split in two; one subset is used for training a supervised Machine Learning (ML) system, and the other subset is used as the test set. The chosen ML was shown not to be critical, as different choices were tried, and the results were consistent across them, proving the robustness of the procedure with the given data set $(h, E, D)$.

\section{Results}

\subsection{Classification of Parkinson's disease through gait data}

As a first example, gait data from healthy subjects and subjects with idiopathic Parkinson's disease was retrieved from the Physionet public database \cite{goldberger00}. The same data has been used in previous studies \cite{frenkel05,yogev05}. The measurement consists of the sum of 8 force sensors in each foot. The subjects walked unassisted for approximately 2 minutes on level ground at their usual, self-selected, comfortable speed. From the measurement, the total force in each foot as a function of time is recorded. Additional data, such as age and gender, were also available. Subjects were stratified into young adults ($<60$ years) and older adults ($>60$ years). The data of 163 subjects was used; the mean age was $65$ with a standard deviation of $9.2$; further details of the studied population can be found in Table \ref{tbl:parkinson}

\begin{table}[!ht]
 \center
 \renewcommand{\arraystretch}{1.5}
{
 \begin{tabular}{cccc}
%&       & \multicolumn{3}{c}{CO} & \multicolumn{3}{c}{PD}\\
& Total & frac. CO & frac. PD \\
\hline
Young & 43  & 0.534 & 0.465 \\
Old   & 120 & 0.425 & 0.575 \\
\hline
 \end{tabular}
}
\caption{\textbf{Gait measurement statistics.} The subjects were divided into two groups: young adults below 60 years, old otherwise. CO refers to the healthy control subjects, while PD refers to Parkinson's patients. frac. refers to the fraction of subjects under each condition.}\label{tbl:parkinson}
\end{table}

\begin{figure*}[!ht]
\centering
\includegraphics*[scale=0.6]{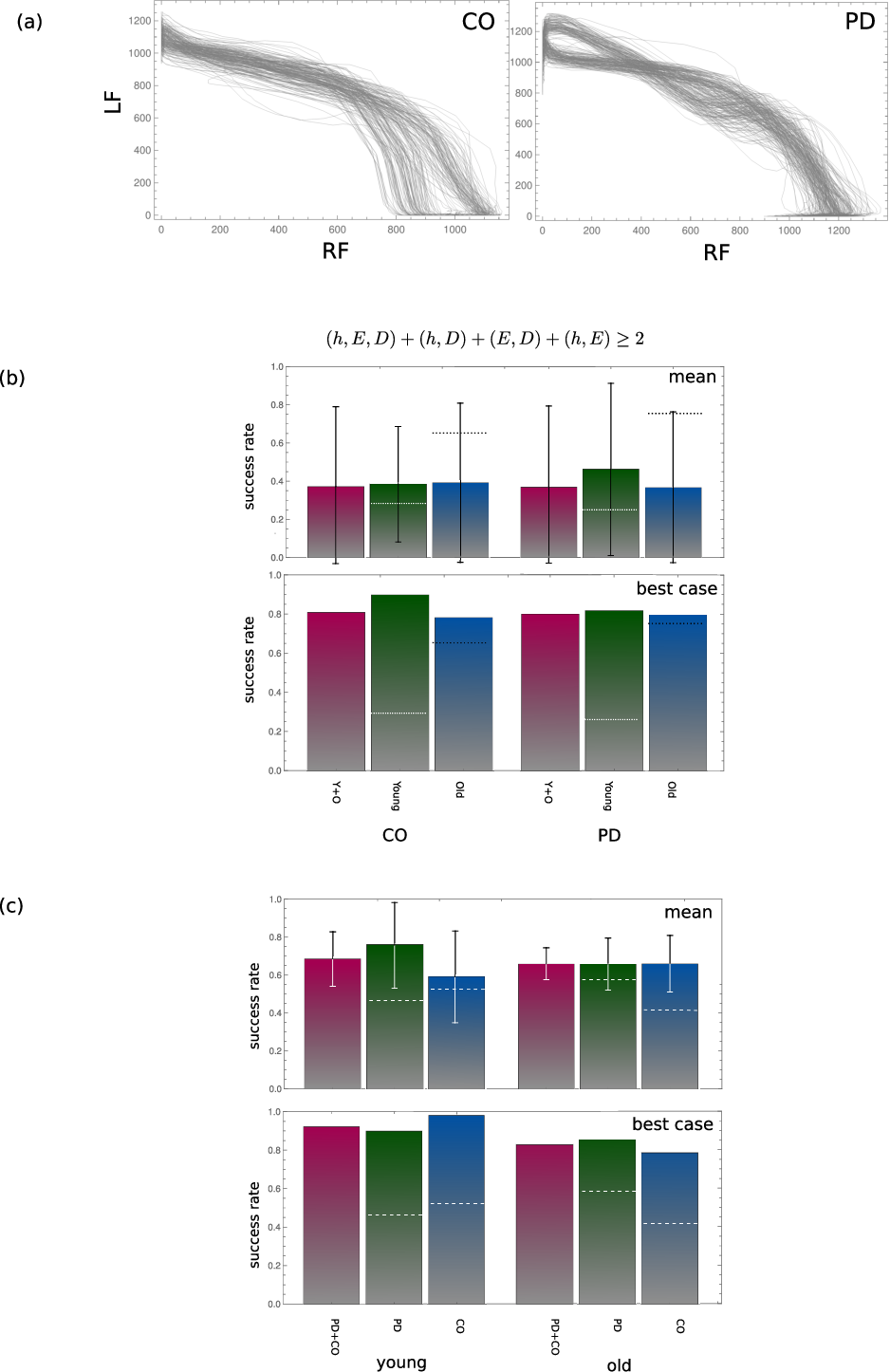}
\caption{\textbf{Parkinson's disease analysis through gait measurements}. Pressure in both feet was recorded as a function of time and taken as a trajectory. (a) Shows the plot of two trajectories, one for a control healthy subject (CO) and one for a subject with Parkinson's disease (PD). $RF$ and $LF$ refers to right and left foot, respectively. Subjects are walking at a normal pace for two minutes. Subject trajectories were discretized using Freeman coding and entropic analysis, and estimating the entropy density ($h$), effective measure complexity ($E$), and fractal dimension ($D$) was performed. From the entropic data, the classification of subjects into young and old was attempted (b). In order to perform the classification, the entropic magnitude was used on a neural network. The subjects were divided into two subsets: one for training, with the same number of young subjects and old patients, and a second subset for the classification trial. This was repeated $100$ times, choosing the training and trial subset randomly. The training set was never larger than $1/3$ of the total number of subjects. Four experiments were carried out with the neural network. In three experiments, two of the three magnitudes were used for training and classification; in a fourth experiment, all three magnitudes were used for training. For each trained neural network, the trial subjects' entropic measures were fed into them for classification. The mean (upper) and best-case (lower) results are presented. The same procedure was followed for classifying the set in CO and PD subjects (c). For the training, the subset was chosen with the same number of healthy and Parkinson's subjects. Subjects classified as PD by two or more trained neural networks were labelled as such. The mean (upper)  and best-case (lower) results are shown. In both (b) and (c), the dashed lines are the fraction of the given subjects for each type.
}\label{fig:parkinson}
\end{figure*}

The data was treated as a trajectory as shown in figure \ref{fig:parkinson}(a). From the trajectory plot, a clear distinctive trend between healthy CO subjects and Parkinson's PD patients could not be found. Following the described procedure, each trajectory was coded into a Freeman string. The grid size was $2.000$.  $(h, E, D)$ was estimated from the coded string. Table \ref{tbl:phed} shows the mean values and standard deviation. The mean entropy density is larger for PD subjects compared to CO subjects in both age groups. Correspondingly, the effective measure complexity $E$ is smaller in the Parkinson's patient with respect to the CO subjects. Changes in the fractal dimension appear less significant than those observed in the entropic magnitudes. While entropy density changes more between CO and PD in the young group, $E$ has the largest variation in the old set. 

\begin{table}[!ht]
 \center
{
 \begin{tabular}{cccc}
& & \text{Young} & \text{Old} \\
\hline
\multirow{3}{*}{CO} & $\left\langle h\right\rangle$  & 0.90(0.09) & 0.93(0.07) \\
                     & $\left\langle E\right\rangle$  & 15.1(0.9) & 15.1(0.8) \\
                     & $\left\langle D\right\rangle$  & 1.04(0.02) & 1.04(0.77) \\
                     \hline
 \multirow{3}{*}{PD} & $\left\langle h\right\rangle$  & 0.97(0.09) & 0.96(0.08) \\
                     &$\left\langle E\right\rangle$  & 14.9( 0.8) & 14.4(1.1) \\
                     &$\left\langle D\right\rangle$  & 1.04(0.01) & 1.04(0.01) \\
 \end{tabular}
}
\caption{\textbf{Mean value of the entropic magnitudes and fractal dimension.} CO corresponds to the control healthy subject. Subjects with Parkinson's disease are labelled as PD. Average entropy density $\langle h \rangle$, effective measure complexity $\langle E\rangle$, and fractal dimension $\langle D\rangle$ were estimated. Subjects were divided into age groups; those over $60$ years were labelled as old, and the rest as young. In parenthesis the standard deviation.}\label{tbl:phed}
\end{table}

The data has two conflicting conditions: the age group and the health condition. Both old age and Parkinson's disease can have the same effect on the gait data, so the first question we try to answer is if stratification in age could be done from the $(h, E, D)$ data. 

We divided the set into two subsets, one for CO and one for PD subjects. As described, in training and the classification stage, the input into the neural network is the tuple $(h, E, D)$. Four neural networks were used; three of them were trained with two of the three magnitudes, namely $(h, E)$, $(h, D)$, $(E,D)$; and a fourth neural network was trained with the whole tuple of three values. A random subset for training in each group, CO and PD, was selected from the set of subjects. The size of the training set was always at most $1/3$ of the total set. The procedure was repeated $100$ times, and training and classification were performed for each. In all cases, no surrogate data was created, and no overlapping between the training set and the trial set was allowed to avoid data leaks and bias. Results for the mean over all instances and the best-case scenario are reported.

Bar plot \ref{fig:parkinson}(b) shows the success rate in age classification on average and in the best case. The average success rate does not show good discrimination, and differences between CO and PD are not significant; although the classification is above the probabilistic case for young people, it is well below for old subjects. The average case is a pessimistic evaluation, training is a supervised procedure, and bad training data can be rejected. The advantage of using the average values for evaluating the success rate is that it avoids, as much as possible with the limited data set, any bias due to overfitting of the neural network. On the other hand, the best scenario case shows a success rate in age discrimination with values above $80\%$ in all cases. The best scenario is an over-optimistic evaluation; results can be driven by overfitting the particular data set and, in the absence of a larger data set, can give a biased evaluation of the classification capabilities. 

As a second question, we ask if classification into CO and PD subjects can be done from the $(h, E, D)$ data. We follow a similar procedure as before, but now, the set is divided into two subsets for young and old subjects. Neural networks are trained following a similar procedure. Bar plots in figure \ref{fig:parkinson}(c) show the classification success rate for the average and the best case. 

In all cases, the classification procedure improved over the probabilistic case (randomly picking a subject and assigning a label). In the mean, improvement over the probabilistic case was the largest for young subjects in classifying the occurrence of Parkinson's, with a $61.5\%$ increase in the detection rate up to $75.1\%$ of the cases. For old subjects, the increase was up to $14.8 \%$, but the success rate for the CO subjects increased in $55.5\%$. The selectivity of the test follows from a balance between the success rate of CO and PD subjects. A too-high success rate for PD patients, with a low success rate for the CO subject, means that the test classifies almost all subjects as Parkinson's patients and the selectivity is low; the same goes for the vice versa case. The overall success rate measures such balance, and we see a success rate of $67\%$ for the young subject, which is an improvement of $34\%$ over the probabilistic case. For the old subjects, the value is relatively close at $66\%$ with an improvement of $32\%$.

The best case scenario shows a much more stark improvement with respect to the probability case as seen in Figure \ref{fig:parkinson}c. For young subjects, the success rate increases by $87.3\%$ for the control group and $95.5\%$ for the Parkinson's patients with respect to the probability case. For the old subjects, the improvement is $85.6\%$ and $49\%$ for CO and PD subjects, respectively.

In this case, the previous comment between using the average or the best-case scenario for evaluating the classification procedure holds equally.

\subsection{Asserting fall occurrence through human posture measurements}

\begin{figure*}[!ht]
\centering
\includegraphics*[scale=0.75]{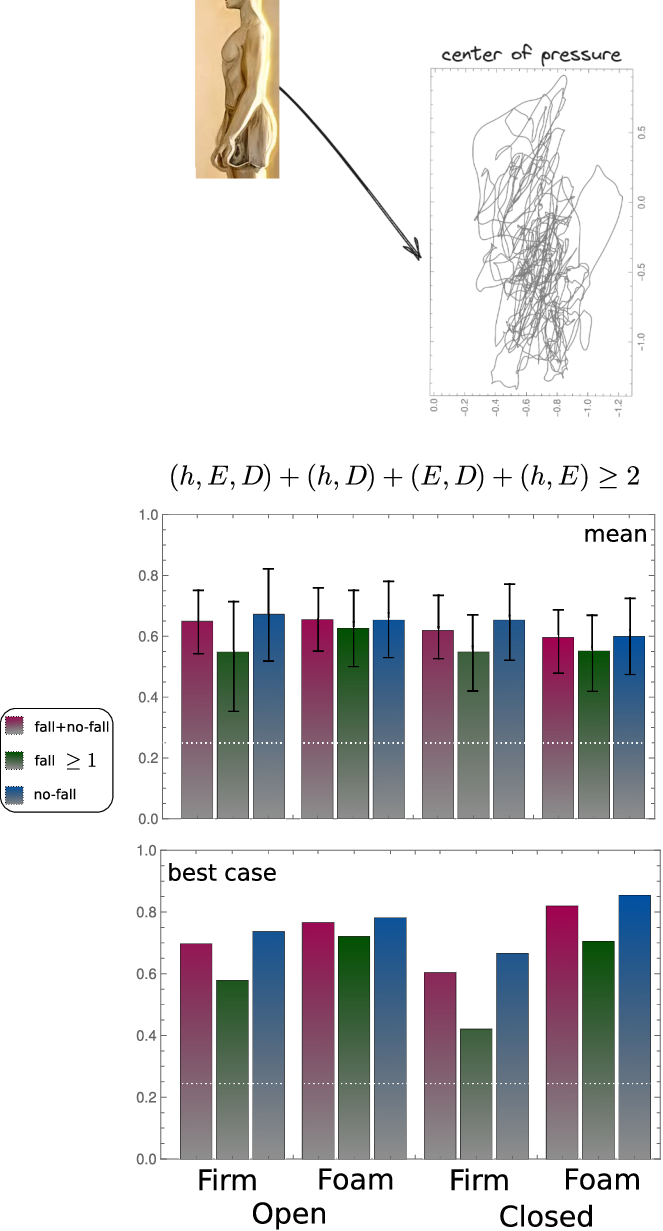}
\caption{\textbf{Fall occurrence through human posture measurements} The centre of pressure of standing subjects with different health backgrounds is recorded for 60 seconds. Four different conditions are set for the experiments. The subject is with its eye open and on a firm surface or a foam soft surface. The same is done with the eyes closed. The measurements result in trajectories, which are coded into a Freeman string. From the character string, the entropy density ($h$), effective measure complexity ($E$) and fractal dimension ($D$) are estimated. The record of falls in the last 12 months is known for each subject. The ability to assert if a subject has fallen from the $(h, E, D)$ measurement is tested. We proceed as in the Parkinson's case. The subject set is split into two non-overlapping sets, one for training and one for trial. As described before, the training set is used to train a   neural network. The trial set is then used to test the classification ability. The middle bar plot is the mean success rate for each condition over the trial set. The dashed line is the fraction of subjects with at least one falling episode. The success rate is measured for the combined fall and non-fall subjects, for the correct determination of the subjects with at least one falling event, and for subjects with no falling event. In all conditions, the success rate is far above the dashed line. The same result is shown for the best case in the classification procedure in the lower bar plot.
}\label{fig:posture}
\end{figure*}

Human balance can be measured in different ways, and its measurement has been used to determine different deficiencies in the servomotor control system. We used data from the publicly available database Physionet \cite{goldberger00}. The measurement is known as static posturography, where the centre of pressure is recorded while the subject is quiet standing (Figure \ref{fig:posture} top). The centre of pressure is the point of application of the vertical ground reaction force vector. It is acquired with a force platform that produces a two-dimensional time series representing the centre of pressure trajectory in the anterior and medial-lateral axes. A set of $76$ participants above $60$ years old were recorded. A detailed description of the experiments can be found in \cite{santos16}. In summary, the subjects were measured while standing still for $60$ seconds in four conditions. Experiments were performed with eyes open and closed, and for each, two surfaces were used, one firm and the other a foam mat. For each condition, three measurements were carried out. The age, health status, and other health status information were gathered. In particular, the number of falling events for each subject was determined in the last $12$ month, from which subjects could be classified as falling and non-falling. Around $1/4$ of the subject has falling events (See table \ref{tbl:post}) 

\begin{table}[!ht]
 \center
 \begin{tabular}{cl|cccc}
       &   & no-fall freq & fall freq \\
 \hline 
 \multirow{2}{*}{Open}  & Firm &  0.750 &  0.25 \\
                        & Foam &  0.753 &  0.246 \\
                        \hline
\multirow{2}{*}{Closed} & Firm &  0.757 &  0.243 \\
                        & Foam &  0.764 &  0.236
\end{tabular}
\caption{\textbf{Human posture experiment statistic.} Four experiments were performed. One set of experiments was done with the eyes open and another set with the eyes closed. For each of these, two experiments were done; in one, the subjects were standing on a firm surface, and in the other, on a soft foam surface. Each subject's record of falling events in the last 12 months was known. Around $1/4$ of the subjects had at least one falling event.}\label{tbl:post}
\end{table}

\begin{table*}[!ht]
\centering
\begin{tabular}{cc|lll|lll}
&&\multicolumn{3}{c}{non-falling}&\multicolumn{3}{c}{falling}\\
 \text{Eyes}&\text{Surface}&\text{$\langle $h$\rangle $} & \text{$\langle $E$\rangle $} & \text{$\langle $D$\rangle $} &\text{$\langle $h$\rangle $} & \text{$\langle $E$\rangle $} & \text{$\langle $D$\rangle $} \\
\hline\\
\multirow{2}{*}{Open} &\text{Firm}& 1.15(0.13) & 11.1(1.4) & 1.05(0.02) & 1.09(0.15) & 11.5(1.5) & 1.04 (0.02)\\
&\text{Foam}& 0.81(0.05) & 16.4(0.7) & 1.01(0.004) & 0.78(0.04) & 16.2(0.7) & 1.01(0.002) \\\\
\multirow{2}{*}{Closed} &\text{Firm}& 1.14(0.14) & 11.1(1.4) & 1.05(0.02) & 1.06(0.12) & 11.7(1.5) & 1.05(0.02) \\
&\text{Foam}& 0.77(0.05) & 16.8(0.8) & 1.01(0.003) & 0.74(0.04) & 17.0(0.7) & 1.01(0.002)\\
\end{tabular}
\caption{\textbf{Mean value of the entropic magnitudes and fractal dimension.} Average Entropy density $\langle h\rangle$, effective measure complexity $\langle E\rangle$, and fractal dimension $\langle D\rangle$ for each experiment type. In parenthesis the standard deviation.}\label{tbl:hedpost}
\end{table*}

Each trajectory was coded into a Freeman string, and the data was processed as in the Parkinson example. The grid size was $2.000$. $(h, E, D)$ was estimated for each experiment type from the coded string. Average values are reported in Table \ref{tbl:hedpost}. $\langle h \rangle$ shows the largest variations between falling and non-falling subjects, where it changes up to $5.5\%$ for the open-eye experiments and $7.5\%$ for the close-eye experiments.

Classification procedures follow the same protocol used in Parkinson's disease studies. Figure \ref{fig:posture} middle and bottom show the results for the mean and best cases, respectively. For all the experiments, classification using the corresponding neural networks outperforms drastically the probabilistic random choice (dashed white lines). 

In both the mean and the best-case scenario, the success rate is generally much better than those achieved in the Parkinson example and above a toss of a coin. In the average results, the success rate can be as high as $62\%$ for the falling subjects in the open-eyes foam-mat surface experiment while keeping a $67\%$ success rate for the non-falling subjects. These results improve for the best scenario case as expected. 

\subsection{Regular and irregular motion in the H\'enon-Heiles model}

As a third example, the orbits of the well-known H\'enon-Heiles potential are studied \cite{henon64}. Conservative Hamiltonian systems of a finite number of degrees of freedom can exhibit regular and irregular behaviour \cite{henon64,percival73}. Both motions can be discriminated by the form of phase space trajectories or by the structure of the frequency spectra of the motion \cite{powell79}. According to KAM theorem \cite{lichtenberg92}, a bound system with analytic Hamiltonian and almost separable has a phase space with significant volumes of regular trajectories or trajectories that behave similarly to those of separable systems and lie in invariant toroids. 

On the other hand, irregular trajectories do not occupy an invariant tori, and they are not conditionally periodic. The regular and irregular regions of phase space are complicated interleaved structures, making rigorous estimates of their volume difficult even for the simplest cases. In certain energy ranges, fixing the energy does not guarantee regular or irregular orbits, as they also depend on the initial conditions.

The H\'enon-Heiles Hamiltonian system is simple in terms of its mathematical description, making it easy to be numerically solved, and it has been extensively described in the literature. The equations of the model follow:

\begin{equation}
\begin{array}{l}
 H(q,p)=\frac{1}{2}(\dot{x}^2+\dot{y}^2)+V(x,y)\\\\
 V(x,y)=\frac{1}{2}(x^2+y^2+2 x^2 y-\frac{2}{3}y^3)\label{eq:hh}
 \end{array}
\end{equation}

Where $V(x,y)$ is the energy term. The two-dimensional potential is almost integrable at the bottom of the well, where the energy is zero, and regular trajectories dominate the phase space. There is an escape energy at $1/6$. As such value is approached, irregular trajectories increase its volume in phase space, becoming significant for an energy value of $1/12$ and dominating near the escape energy \cite{henon64,percival73,powell79}. 

\begin{figure*}[!ht]
\centering
\includegraphics*[scale=0.9]{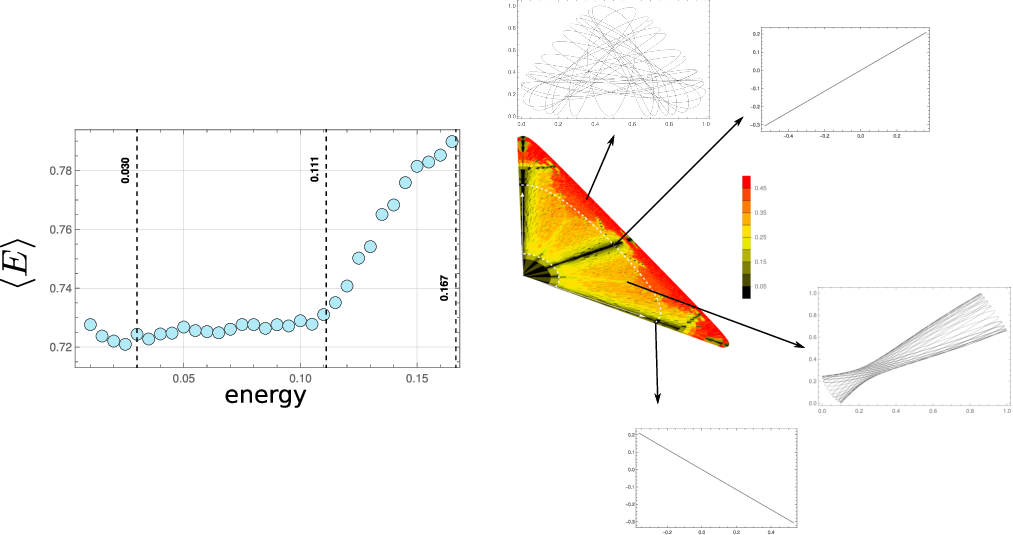}
\caption{\textbf{Regular and irregular orbits in the H\'enon-Heiles potential.} \textbf{Left:} Effective measure complexity analysis of orbits in the H\'enon-Heiles potential. $\langle E \rangle$ is the mean value of the effective measure complexity. At low energies, regular orbits are almost the total of all orbits; after a transient, at roughly $0.030$ the effective measure complexity settles in a plateau in the interval $[0.030, 0.111]$ at a mean value of $0.726\pm 0.002$ where almost all orbits are still regular. For energy values larger than $0.111$, the mean entropy density increases monotonically until an energy of $1/6$, which corresponds to the escape energy. This the region where irregular  orbits starts to dominate. \textbf{Right:} In the inner energy well, below $1/6$ there is a wealth of different orbits depending on the initial conditions. Taking initial starting from the rest condition and different starting position, the entropy density was calculated from the chain coded trajectories and shown as a color label at each point in the well. The inner contour (in white) corresponds to an energy value of $0.01$. For energies below this value, all trajectories show trivial behavior from the predictability point of view. In the intermediate region between both contour lines, corresponding to the plateau, unpredictability increase but still with values of $h$ below $0.35$, the trajectories are more creative while still being regular. For larger energy values, irregular orbits start to dominate shown as the red regions in the plot. In certain symmetric directions, orbits are periodic and the corresponding entropy density is near zero almost in the whole energy range, this can be seen as straight black lines.
}\label{fig:hh}
\end{figure*}

There have been several methods for determining regular and irregular trajectories \cite{henon64,powell79} and the fraction of both types of orbits in phase space as a function of energy. Instead of repeating such results, here we follow the evolution of the mean value of the effective measure complexity $E$  of the orbit trajectory as a function of energy (Figure \ref{fig:hh}-left). In such a way, we are looking at how structured, in the mean, the orbits appear for a given energy. The mean effective measure complexity for an energy value of $0.01$ is around $0.728$; from there, it slightly decreases with increasing energy to reach a plateau at an energy of $0.030$ with a value of $0.723$. $\langle E \rangle$ starts again to increase significantly at an energy value of $0.110$ up to $0.165$, reaching a value of $\langle E \rangle =0.790$. This last value of energy is on the edge of the escape energy. For larger energy values, the number of bounded orbits falls dramatically, making the reported mean values not indicative of the system's behaviour.

The question of what to consider irregular is one with several answers, none conclusive and all informative. Instead of looking for another criterion for irregularity, calculating $\langle E \rangle$ allows us to focus quantitatively on unpredictability. If we agree that all trajectories are regular at an energy of $0.01$, then there is a mean value of $0.724$ of structuring in such trajectories for such low energy values. The question is, then, how much pattern can regular trajectories accommodate? The answer seems to be in the analysis of the plateau region between energies $0.030$ and $0.100$. This region roughly corresponds to the reported interval where regular trajectories are almost the totality. The mean value of $\langle E \rangle$  in such region is $0.726\pm 0.002$. As the energy goes above the plateau, the entropy density increases with a linear slope of $1.13$ E/energy. A similar analysis could be made with the entropy density.

Finally the dependence of the orbit type with initial condition was explored. We proceed as follow. The initial momentum of the trajectories were taken as zero, the initial position was taken at points within the central well (energy below $1/6$ ) of the H\'enon-Heiles potential, the corresponding orbit was calculated and the entropy density of the trajectory computed after chain-coded. The color map in Figure \ref{fig:hh}-right shows the value of $h$ at each initial point. Only one third of the whole circle is shown, between $-\pi/3$ and $\pi/2$, due to the symmetry of the potential. As the point move from the centre, the energy of the system increase. Below $0.01$, in the inner contour, all orbits are of low entropy density, near zero. Trajectories are not only regular but rather trivial in terms of predictability. Here we are well into the KAM region, where trajectories comes from an almost integrable system and lie in an invariant toroid. Within the outer contour line, entropy density is below $0.35$, the system is still mainly in the KAM  regime but the unpredictibility has increase, trajectories, yet regular, can be more creative. At energies around the escape energy, outside the outer countour are mostly irregular and do not lie in the KAM torus. Another interesting feature is that certain directions, namely at angles $-\pi/6$, $\pi/3$ and $\pi/2$, exhibit almost zero entropy density, even at high energies. This orbits in this directions are linear, going backward and forward. Increase in energy merely increases the amplitude of the periodic orbit. 

\section{Conclusions}

The results describe an effective procedure for the entropic analysis of trajectories, allowing the definition of a trajectory entropy density and related quantities. The key point is to code any continuous curve into a code-chain sequence over a finite alphabet, where we used the well-established Freeman coding of shape. Information production, through entropy density, and redundancy or pattern production, through the effective measure complexity can then be estimated. Furthermore, it is straightforward to calculate the fractal dimension of the curve from the Freeman code. 

The robust nature of the method allows it to be applied to a wide range of subjects. For experimental physiological data of gait and human posture, the extracted quantities of $h$, $E$ and $D$ from the pertinent trajectories, can be used to classify subjects according to Parkinson's disease, in the gait experiment case, and the occurrence of falls, in the human posture data. 

A further application was explored in analysing regular and irregular orbits in the well-known H\'enon-Heiles dynamic equations. Irregular motion increasingly happens as energy goes beyond a threshold value. Our analysis allows us to estimate how much disorder regular motion can accommodate with respect to a fully periodic orbit as quantified by the effective measure complexity or the entropy density in the plateau region. Beyond the plateau region, the increase of the mean effectve complexity is associated with an increasing fraction of irregular orbits; in this case, the rate of increase can be determined, and the maximum value of $E$ before the escape energy is reached can be estimated. The entropy density map of the central well, shows the dependence of the trajectories with the initial condition. 

\section{Methods}

\begin{enumerate}

\item \textbf{Freeman coding}. In the H\'enon-Heiles analysis, each trajectory was coded with three rotations of $\pi/3$ to avoid any dependence on the orientation with respect to the grid. The orientation with the lowest entropy density was chosen. 

\item \textbf{Lempel-Ziv factorization.} Consider for a  sequence $f=f_1 f_2\dots f_N$. A factorization $E(f)=f(1,h_1)f(h_1+1,h2)\dots f(h_{m-1}+1,N)$, where $f(i,j)$ is the substring $f_i f_{i+1}\dots f_j$, and each symbol $f_i$ is drawn from the Freeman alphabet $\Sigma={0, \ldots 7}$ of cardinality $8$. $E(f)$ is called an exhaustive history of the sequence $f$, if any factor $f(h_{j-1}+1, h_j)$ is not a substring of the string $f(1,h_j-1)$, while $f(h_{j-1}+1, h_{j}-1)$ is. The LZ76 complexity $C(f)$ is the number of factors of the exhaustive history $E(f)$. For example, the exhaustive history of the sequence $u=1101000011001$ is $E(f)=1.10.100.0001.110.01$, where a dot separates each factor, and $C(f)=6$.

In general, $C(f)$ for a length $N$ string, is bounded by\cite{lz76}
\[
 C(f) < \frac{N}{(1-\varepsilon_N)\log_{\sigma}{N}},
\]
where
\[
 \varepsilon_{N}=2\frac{1+\log_{\sigma}{\log_{\sigma} \sigma N}}{\log_{\sigma}N}.
\]
$\varepsilon_{N}$ is a slowly decaying function of $N$, leading to an asymptotic value
\[
 C(f) < \frac{N}{\log{N}},
\]
for large enough $N$.

Ziv\cite{ziv78} proved that if $f$ is the infinite length output from an ergodic source with entropy rate $h$, then
\[
\limsup_{N\rightarrow\infty}\frac{C[u(1,N)]}{N/\log{N}}=h
\]
almost surely. This is the basis for estimating the entropy density using the Lempel-Ziv factorization.

\item For the classification procedures, the Classify function as implemented in Wolfram Mathematica computation system \cite{Mathematica} was used. The procedure used the ''LogisticRegression'' method \cite{murphy12} as the machine learning method, using as optimization a limited memory Broyden-Fletcher-Goldfarb-Shanno algorithm \cite{liu89} and regularization term in the square of the parameter matrix ($\lambda_1=0$, $\lambda_2>0$).

\item In the H\'enon-Heiles model, the non-linear differential equations (\ref{eq:hh}) were numerically solved using an explicit Runge-Kutta method with adaptive embedded pairs with a maximum step size of $0.001$ as implemented in Wolfram Mathematica computation system \cite{Mathematica}. For each energy, a total number of $1400$ orbits was calculated. Orbits that escaped the central well of the H\'enon-Heiles potential defined by the triangle vertices $(0,1),(-\sqrt{3}/2,-1/2),(\sqrt{3}/2,-1/2)$ were discarded. For energies $<1/6$, only a few (at most $7$) orbits were discarded at each energy value. Before Freeman coding, trajectories were scaled to a $1\times 1$ square area. The grid resolution was set at $0.05$.

\item The used physiological data can be found in the open repository at https://physionet.org/about/database/

\item \textbf{Orbit calculation.} The H\'enon-Heiles differential equations  were solved using the built-in Runge-Kutta procedure in Wolfram Mathematica computation system \cite{Mathematica} with maximum step size of 0.001 (The Mathematica notebook can be found in the Supporting Information).
\end{enumerate}

%**********************************************************************
\section{Acknowledgments}
%**********************************************************************
CITMA is acknowledged for financial support under project CARDENT, grant PN223LH010-053. EER would like to thank the AvH Foundation for direct support under a visiting scholarship grant. K\'arel Garc\'ia is acknowledged for valuable discussions. The University of Havana is acknowledged for its computer support and working environment. Ra\'ul Izquierdo is thanked for his help in the design of the figures.

% Bibliography
%\bibliography{entropyfreeman}

\end{document}